# High Q-factor diamond optomechanical resonators with silicon vacancy centers at millikelvin temperatures

Graham D. Joe[1,†], Cleaven Chia[1,2,†], Benjamin Pingault[1,3,4], Michael Haas[1], Michelle Chalupnik[1], Eliza Cornell[1], Kazuhiro Kuruma[1], Bartholomeus Machielse[5], Neil Sinclair[1], Srujan Meesala[6], Marko Lončar[1]*

[1] *John A. Paulson School of Engineering and Applied Sciences, Harvard University, 29 Oxford Street, Cambridge, Massachusetts 02138, USA*
[2] *Institute of Materials Research and Engineering (IMRE), Agency for Science, Technology and Research (A\*STAR), 2 Fusionopolis Way, #08-03 Innovis, Singapore 138634, Republic of Singapore*
[3] *QuTech, Delft University of Technology, 2600 GA Delft, Netherlands*
[4] *Center for Molecular Engineering and Materials Science Division, Argonne National Laboratory, Lemont, IL 60439, USA*
[5] *Department of Physics, Harvard University, Cambridge, MA 02138, USA*
[6] *Institute for Quantum Information and Matter and Thomas J. Watson, Sr., Laboratory of Applied Physics, California Institute of Technology, Pasadena, California 91125, USA*
*† These authors contributed equally to this work*
*\* Corresponding author. Email: loncar@seas.harvard.edu*

Phonons are envisioned as coherent intermediaries between different types of quantum systems. Engineered nanoscale devices such as optomechanical crystals (OMCs) provide a platform to utilize phonons as quantum information carriers. Here we demonstrate OMCs in diamond designed for strong interactions between phonons and a silicon vacancy (SiV) spin. Using optical measurements at millikelvin temperatures, we measure a linewidth of 13 kHz (Q-factor of ~$4.4\times10^5$) for 6 GHz acoustic modes, a record for diamond in the GHz frequency range and within an order of magnitude of state-of-the-art linewidths for OMCs in silicon. We investigate SiV optical and spin properties in these devices and outline a path towards a coherent spin-phonon interface.

## 1. Introduction

Micro/nanomechanical systems have emerged as a promising platform for quantum science and technology owing to their ability to coherently interact with a wide variety of quantum systems [1]–[3]. Optomechanical crystals (OMCs), which confine optical photons and acoustic phonons on a wavelength scale and thus can enable efficient photon-phonon interactions [4], have attracted significant attention. The optomechanical interaction has been used to demonstrate quantum ground-state cooling of a macroscopic mechanical mode [5], squeezed light [6], microwave-to-optical transduction [7], and a telecom spin-photon interface using an intermediary mechanical mode [8], as depicted in Fig. 1(a). Wavelength-scale confinement of phonons in OMCs could also allow for strong coupling between a mechanical mode and a strain-sensitive defect spin qubit [9]. Single-crystal diamond is a natural platform for such devices since it hosts a variety of color-center spin qubits, and features low mechanical dissipation, large Young's modulus, and a wide optical transparency window [10]–[12]. This has motivated activity on coherent acoustic driving of nitrogen vacancy (NV) spins in a variety of diamond-based devices such as bulk-acoustic resonators and cantilevers [2], [13]–[15]. Relative to the NV center, the silicon vacancy center (SiV) in diamond is better suited as a spin-phonon interface since it provides nearly four orders of magnitude higher strain susceptibility (~100 THz/strain) for the ground state spin [9], [16]. Recently, this has been leveraged to demonstrate low power and coherent acoustic control of a single SiV spin, and proximal nuclear spins [3], [17]. Importantly, the SiV maintains good optical properties in nanofabricated structures in terms of its narrow zero-phonon line (ZPL) and spectral stability, which allow for high fidelity and high efficiency spin-photon entanglement generation [18]. OMCs in diamond have been demonstrated and proposed to enhance the spin-phonon interaction strength towards the strong spin-phonon coupling regime [19], [20]. However, their performance at millikelvin temperatures and their compatibility with SiV centers have not yet been investigated. Operation in this temperature regime is crucial to maintain good SiV spin coherence, improve mechanical quality factors, and thereby, achieve high cooperativity spin-phonon interactions capable of operating near the single-phonon regime.

In this work, we report diamond OMCs which support optical and mechanical modes measured at $f_{opt}$ = *189.6* THz (1580.8 nm) and $f_{mech}$ = *5.76* GHz with high quality factors of $Q_{opt}$~*15,000* and $Q_{mech}$~*440,000* respectively, measured at millikelvin temperatures. These OMCs have a large simulated zero point mechanical fluctuation of $x_{zpf}$ = *3.58* fm and an expected spin-phonon coupling strength ($g_{sp}$) of up to 1.65 MHz, which is large relative to our measured mechanical linewidth (13±1 kHz) and previously measured SiV spin decoherence rates (~100Hz) [21]. The diamond OMCs were fabricated with an angled etching technique [22]. By measuring the dependence of the mechanical linewidth on the optical pump power, we inferred an optomechanical coupling strength of $g_{OM}/2\pi$~*43.5 ± 0.2* kHz at the single photon and phonon level. We characterized single SiV centers embedded in these high-Q diamond OMCs and observed excellent optical properties. However, we observed a reduced spin lifetime ($T_1$) of SiV centers in OMCs compared with expected values [21], likely due to optically-induced local heating. Such heating, caused by parasitic optical absorption, has previously been noted as a source of excess acoustic damping in OMCs at millikelvin temperatures [23], [24]. We discuss alternative OMC device architectures to mitigate the impact of optically-induced heating on spin and mechanical properties.

## 2. Cavity design and fabrication

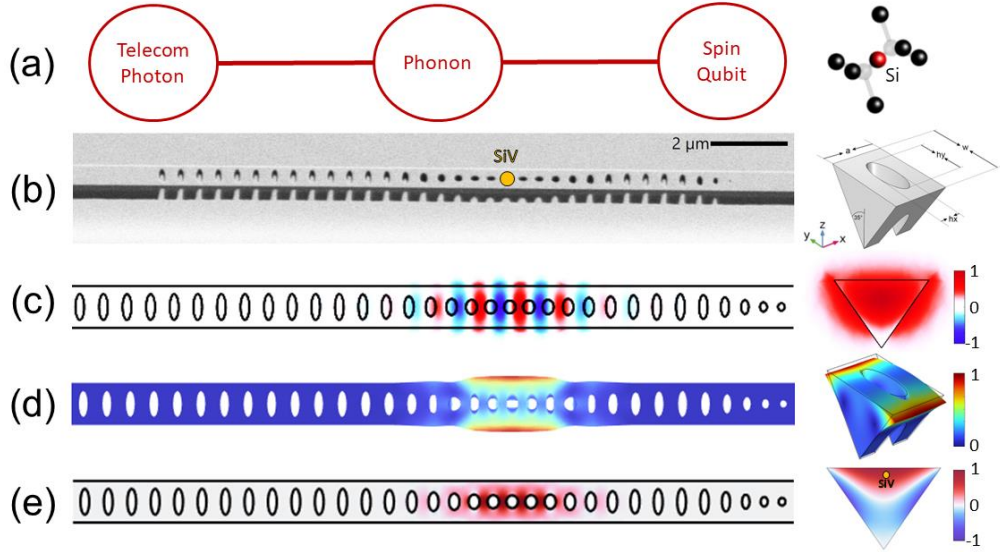

Fig. 1. (a) Illustration of the coupled optical, mechanical, and spin degrees of freedom in our SiV-OMC devices. Inset shows the atomic structure of the SiV center in diamond. (b) Scanning electron micrograph of a typical diamond OMC device fabricated by an angle etching technique. The location of the SiV center is indicated with an orange dot. Inset shows the schematic of a unit cell of diamond optomechanical crystals with geometric parameters of $(a, w, h_x, h_y) = (528, 936, 197, 578)$ nm and a bottom apex half-angle $\theta = 35°$. (c) Optical mode profile ($E_y$ component) of fundamental quasi-transverse electric (TE) mode at $f_{opt} = 197.2\ THz$ ($\lambda_{opt} = 1,520\ nm$). Inset shows the cross-sectional profile of the optical mode. (d) Displacement and (e) strain profile of the mechanical (flapping) mode supported by the OMC device at $f_{mech} = 6.6\ GHz$. The insets of (d) and (e) show the 3D displacement profile of a unit cell and the cross-sectional strain profile, respectively.

Our diamond OMCs consist of a free-standing diamond waveguide with triangular cross-section [22], and a one-dimensional (1D) array of elliptical air-holes [Fig. 1(b)]. To form photonic and phononic cavities in the same region, we locally vary the lattice constant and eccentricity of the air-holes. Details of the tapering method can be found in the supplementary material. The optical and mechanical modes of diamond OMCs are then simulated using the finite-difference time-domain (FDTD) method (Lumerical) and the finite element method (FEM) (COMSOL). Fig. 1(c) shows the field distribution of a quasi-transverse electric (TE) optical resonance at $\lambda_{opt} = 1,520\ nm$. The calculated $Q_{opt}$ and mode volume are $5 \times 10^6$ and $0.50(\lambda_{opt}/n)^3$, respectively, where $n = 2.4$ is the refractive index of diamond. The designed OMC also supports a so-called "flapping" mode [Fig. 1(d)], with a large zero-point motion $x_{zpf}$ of 3.58 fm at 6.6 GHz. This mode has large mechanical displacements near the edge of the cavity region, and large strain fields near the center [Fig. 1(e)], making it possible to achieve large optomechanical coupling to a quasi-TE optical mode [19]. The maximum strain amplitude, and optimal SiV spin location, is in the center of the waveguide between the holes of the cavity region and close to the top surface of the device. The details of the cavity simulation can be found in the supplementary material.

Given the calculated profiles of optical and mechanical modes, we estimate the optomechanical coupling rate between the optical mode and mechanical flapping mode

$g_{OM}/2\pi$ to be 127 kHz. This value is comparable with the values previously reported in diamond OMCs at similar frequencies [19]. We also calculated the spin-phonon coupling rate $g_{SP}$ from the simulated mechanical flapping mode assuming a single SiV is coupled to the maximum strain located at depth of 20 nm below the diamond surface [see inset in Fig. 1(d)]. The maximum achievable $g_{SP}/2\pi$ at this implantation depth is 1.65 MHz.

The OMCs are fabricated on electronic-grade single-crystal diamond substrates (Element Six). The OMC fabrication process, previously described in [25], consists of the following steps: (i) patterning of alignment markers, (ii) generation of SiV centers at a 20 nm depth via masked implantation and annealing [26], [27], (iii) OMC fabrication via vertical and angled etching [22]. Steps (ii) and (iii) are performed aligned to the markers defined in (i) [see supplementary material for more details of the fabrication process]. For this sample, silicon ions are implanted in the diamond substrate with a dose of $1.25 \times 10^{12}$/cm$^2$ and beam energy of 27.5 keV. The sample is then annealed at ~1400K in an ultrahigh vacuum furnace at pressures $< 10^{-6}$ Torr to activate SiV centers. Fig 2(a) shows a scanning electron microscope (SEM) image of the fabricated OMCs. The structure incorporates a tapered waveguide region at its end (right side) for high optical coupling efficiency to a tapered optical fiber [28]. To mechanically support the OMC and thermally anchor it to the cold bath of the diamond substrate for cryogenic experiments, we introduced two adiabatically widened anchors, designed to have low scattering loss, between the OMC and the tapered waveguide, and one on the other side between the OMC and the clamp [Fig. 3(a)].

## 3. Diamond optomechanical crystal characterization

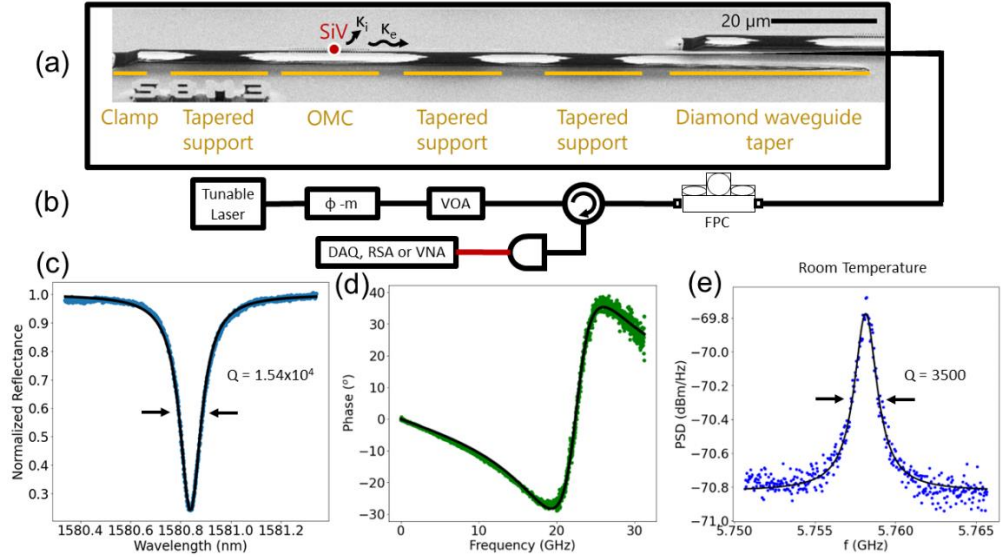

Fig. 2. Room temperature characterization of the diamond OMC. (a) SEM of a diamond OMC device with a tapered waveguide for adiabatic coupling with a tapered optical fiber. Our optical cavities are designed to be single-side waveguide coupled with an extrinsic coupling rate $\kappa_e$ and an intrinsic optical loss rate $\kappa_i$. (b) Simplified schematic of the characterization setup. Φ-m: electro-optic phase modulator, VOA: variable optical attenuator, DAQ: data acquisition device, RSA: real-time spectrum analyzer, VNA: vector network analyzer. (c) Normalized reflection spectrum of the OMC optical mode. A Lorentzian fit (black curve) yields $\lambda_{opt} = 1580.8$ nm and a linewidth of 12.3

GHz (total optical Q factor of 1.54×10⁴). (d) Phase response of the optical resonance. $\kappa_e/(2\pi) \sim 9\ GHz$ is obtained by the fitting (black curve). (e) Power spectral density (PSD) of the mechanical "flapping" mode measured on the RSA with a laser detuning of $\Delta = -\kappa/2$. A Lorentzian fit yields a mechanical linewidth of 1.63 MHz (mechanical Q factor of 3500).

We performed room temperature characterization of the fabricated OMCs using a home-built optical fiber coupling setup [19], illustrated in Fig. 2(b). We investigate the optical mode of the OMC by measuring the reflection spectrum [Fig. 2(c)] with a ~1550 nm tunable laser source coupled via the tapered waveguide - tapered optical fiber interface. We fit the resonance with a Lorentzian function and obtained a linewidth of $\kappa/2\pi = 12.34$ GHz (corresponding to an optical Q of 15,400). To determine whether the cavity Q is limited by intrinsic or extrinsic losses, we measured the phase response of the cavity mode [29]. This is accomplished by detuning the probe laser from the cavity resonance by 0.18 nm (21.3 GHz), then sweeping a sideband generated by an electro-optic (EO) phase modulator across the cavity resonance. The resulting RF beat note of the reflected optical power is recorded on a high bandwidth photodiode connected to a vector network analyzer (VNA). Fig 2(d) shows the phase response of the optical resonance. Using $\kappa$ extracted from the cavity reflection spectrum, a fit to the phase response allows us to extract intrinsic and extrinsic cavity loss rates $(\kappa_i, \kappa_e)/2\pi = (3.17, 9.17)$ GHz where $\kappa = \kappa_i + \kappa_e$. We can therefore conclude that our optical cavity is overcoupled, deviating somewhat from the targeted critical coupling condition (see supplementary material). For mechanical mode spectroscopy, we set the probe laser detuning from the cavity resonance ($\Delta$) to $\pm\kappa/2$ where displacement sensitivity is maximum, we then measured the RF modulation of the probe laser due to the thermal occupation of the mechanical mode [30]. The reflected signal is detected by a high bandwidth photodetector connected to a real-time spectrum analyzer [Fig 2(b)]. Fig. 2(e) shows the power spectral density (PSD) of the "flapping" mechanical mode measured at room temperature. By fitting the PSD to a Lorentzian function, a room temperature mechanical quality factor of $Q_{mech} \sim 3,500$ is measured (in air), likely limited by multi-phonon damping mechanisms [31]. Given the measured optical cavity loss rate ($\kappa = 12.34$ GHz) and the frequency of the targeted mechanical mode ($f_{mech} \sim 5.76$ GHz), our OMC system is in the sideband unresolved regime, where a probe laser detuning of $\Delta = \pm\kappa/2$ is optimal [32]. Future improvements in fabrication may enable us to fabricate sideband resolved resonators with large sideband asymmetries.

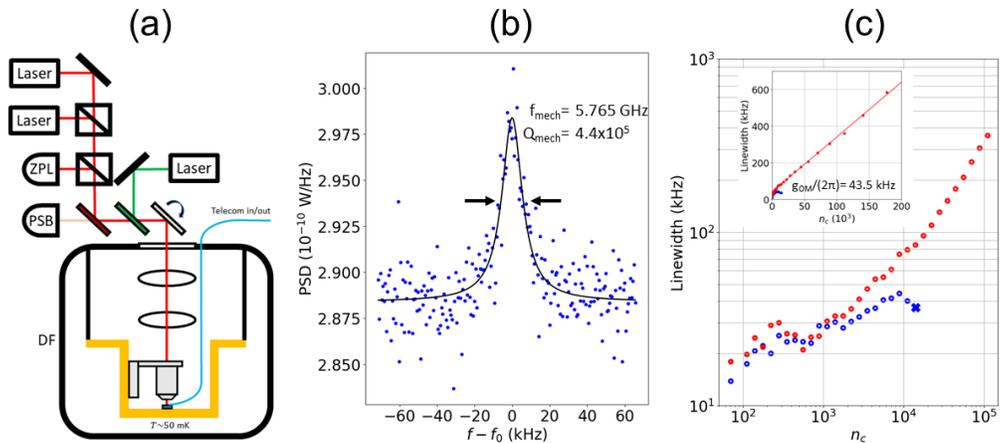

Fig. 3. Millikelvin temperature characterization of diamond OMCs. (a) Schematic of the dilution refrigerator setup and optical path. SiV centers are measured with a confocal microscope built on top of the cryostat. A single-ended tapered optical fiber is used to couple telecom wavelength light into the OMC device of interest. Two lasers are used to address the SiV spin-selective optical transitions, and a pulsed green laser is used as a charge repump for the SiV. Zero-phonon line (ZPL) and phonon sideband (PSB) emission are collected on separate avalanche photodiodes. (b) Mechanical mode spectrum of an OMC device measured at ~50mK with the probe laser parked on the blue sideband ($\Delta = \kappa/2$) at low input power ($n_c \sim 100$). The mechanical linewidth is 13±1 kHz, corresponding to a mechanical Q factor of 4.4×10$^5$ (c) Mechanical linewidth of the OMC device as a function of intracavity photon number $n_c$. Blue (red) points are the experimental data taken with laser tuned to the blue (red) optomechanical sideband of the cavity ($\Delta = \pm\kappa/2$). The blue x at high $n_c$ on the blue sideband corresponds with the OMC's mechanical lasing threshold. Inset shows the corresponding linear plot. A fit to the linear portion of the red sideband (red solid line) yields $g_{OM}/2\pi \sim 43.5 \pm 0.2\ kHz$.

Next, we characterized the diamond OMCs at millikelvin temperatures inside a dilution refrigerator (Bluefors LD250). The diamond sample is mounted to a sample assembly which is thermally connected to the mixing plate at a base temperature of ~50 mK. The optical and mechanical resonances of the OMC are measured via a tapered optical fiber routed into the refrigerator. Fig. 3(b) shows the measured PSD at ~50 mK of the same device and mechanical mode shown in Fig. 2(c). The selected spectrum was taken with the laser tuned to the blue optomechanical sideband ($\Delta = \kappa/2$) and at a power level corresponding to an intra-cavity photon number of $n_c \sim 100$. We observe mechanical linewidths as low as 13±1 kHz, corresponding to $Q_{mech} \sim 440,000$, which is a record high value for diamond OMCs. As expected, the Q factor is dramatically improved at millikelvin temperatures compared to that at room temperature due to the lack of multi-phonon damping mechanisms at such a low temperature. We also investigated the laser power dependence of this mechanical linewidth when the laser is red and blue detuned from the optical resonance by $\pm\kappa/2$ in Fig. 3(c). On the red sideband, the mechanical linewidth broadens with increasing optical power. On the blue sideband, at low pump powers ($\sim n_c < 9 \times 10^3$), the mechanical linewidth also tends to increase with increasing power, which deviates from the predicted decrease in linewidth if only optomechanical backaction [5] is considered. This increase in the linewidth can be attributed to optically-induced heating which is independent of the sign of the laser detuning with respect to the optical resonance [33]. At high pump powers ($\sim n_c > 9 \times 10^3$) on the blue sideband, the mechanical linewidth saturates to an upper limit and then begins to decrease until eventually reaching mechanical lasing, indicated by the blue x in Fig. 3(c) (see also the integrated mechanical spectra in the supplementary material). This suggests that at high optical powers, optomechanical back-action exceeds the thermally broadened mechanical linewidth [34]. Therefore, to extract the optomechanical coupling rate $g_{OM}$, we performed a linear fit to the mechanical linewidth vs. $n_c$ data on the red sideband in the high optical power region above $n_c \sim 9 \times 10^3$. As shown in the inset of Fig. 3(c), this allows us to obtain an upper bound on the optomechanical coupling rate $g_{OM}/2\pi \sim 43.5 \pm 0.2$ kHz at the single photon and phonon level [30]. The difference between the measured and simulated (127.4 kHz) values could be due to deviations, asymmetries, and disorder in the fabricated device geometry with respect to the nominal design.

## 4. Silicon vacancy spectroscopy

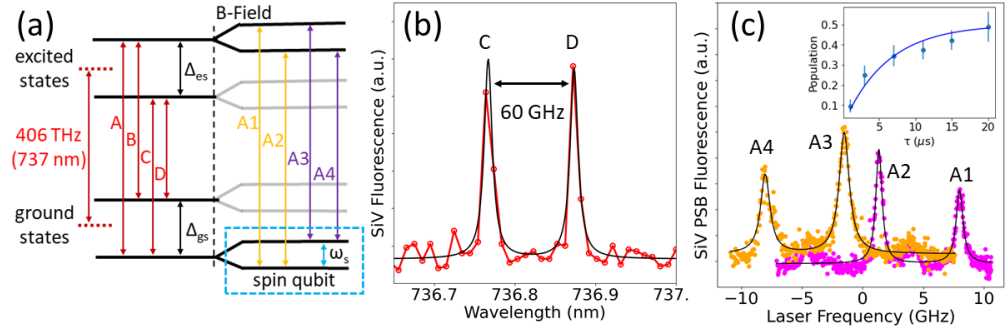

Fig. 4. (a) Energy level diagram of the relevant SiV states. (b) Optical spectrum of the zero magnetic field SiV ZPL resonance fluorescence when pumping the A-line with 500 nW of power at the sample. (c) Photoluminescence excitation (PLE) spectrum of the SiV A-line spin-selective optical transitions in the presence of a 0.4 T z-axis magnetic field (misaligned from the SiV axis by 54.7°), average optical linewidth is 815 MHz. Laser frequency is defined relative to the zero magnetic field A-line transition frequency. Inset shows the population recovery of the SiV spin qubit measured with a pump-probe technique on the A2 and A3 transitions, the decay time constant T1 is 6.1±0.7 $\mu$s. The blue curve is a fit result based on an exponential function.

Finally, we performed optical spectroscopy for single SiV centers embedded in the same OMC device shown in Fig. 2 and Fig. 3 using a home-built confocal microscope setup with an objective lens (Attocube LT-APO-VISIR, NA=0.82) incorporated into the dilution refrigerator [Fig. 3(a)]. Fig. 4(a) shows the SiV energy level structure. There are two orbital levels in the ground state and two in the excited state resulting in four optical transitions labeled A~D. Fig. 4(b) shows the resonance fluorescence spectrum of an SiV center when a laser is resonantly exciting the A transition. We clearly observed emission in the corresponding C and D transitions originating from the same SiV center. From the distance between C and D transition lines, we infer a ground state splitting $\Delta_{gs} \sim 60$ GHz, which is slightly larger than the zero-strain value of ~46 GHz [9]. This indicates that the SiV is slightly strained, likely due to residual strain induced by ion implantation or nanostructuring [9], [16]. Low-strain SiV centers are advantageous for efficiently coupling to mechanical motion since the strain susceptibility of the spin transition decreases with increasing strain [3], [9]. To perform spin spectroscopy on the same SiV center, we then applied a 0.4 T magnetic field along the lab z-axis (54.7° from the SiV axis), which further splits each energy level in two due to the spin ½ of the SiV. This in turn splits each optical transition in four. To carry out photoluminescence excitation (PLE) measurements of SiV centers, we scanned a tunable laser over the zero-phonon line (ZPL) transitions [named A~D in Fig. 4(a)] using a tunable CW Ti:Sapphire laser (M Squared). We also used a 520 nm green laser for periodically re-pumping the sample to maintain the charge state of the SiV [16]. Fig 4(c) displays the high-resolution PLE spectrum measured for the A transition, showing four peaks (named A1~A4) resulting from Zeeman splitting of the A transition under magnetic field. In these measurements, we used a second tunable CW laser to repump one of the spin-dependent optical transitions and increase the visibility of other transitions. PLE resonances from A1 (A3) and A2 (A4) transitions are clearly observed by scanning the first CW laser while the stationary repump laser is resonant with the A3 (A2) transition. To examine the spin properties of the SiV centers, we conduct spin lifetime (T1) measurements on A3 using a pump-probe technique [21], consisting of initialization and readout pulses with a variable inter-pulse time delay, while a repump pulsed laser is tuned to the A2 transition (see details in the supplementary material). The inset of Fig. 4(c) shows the population recovery from the initialized spin state to the other spin state as a function of the inter-pulse delay time. The observed T1 is 6.1±0.7 $\mu$s, which is much shorter than the expected value of milliseconds at millikelvin temperatures [21]. A likely reason for the reduced spin T1 is confocal laser-induced heating of the local environment of the SiV center. Such heating can lead to elevated thermal occupation of the high frequency (~50 GHz) phonon modes responsible for incoherent transitions between the ground state orbital levels, and thereby a reduction in spin T1 [9]. Note that telecom laser power input via the tapered waveguide led to an unexpected SiV fluorescence quenching [8] (see supplementary material). To mitigate the effects of laser-induced heating on SiV centers, it is essential to reduce the necessary optical probe power required for SiV control and read-out. Indeed, cavity QED experiments with SiV centers [26] accomplish this by embedding SiV centers in photonic crystal cavities which support resonances close to SiV ZPL transitions (737 nm), and delivering probe signals via a tapered optical fiber interface. This reduces the laser probe powers to sub-nW levels [26], which is orders of magnitude smaller than the power directed to SiV centers in our experiments (~1 $\mu W$). We note that this was not possible in our work where the optical cavity is designed to operate at telecommunication wavelengths, far detuned from the SiV center's ZPL. As a result, for the devices used in this work, optical excitation and collection of SiV fluorescence through the diamond waveguide port is less efficient than it is through the confocal microscope. Future work will focus on realizing diamond OMCs with optical resonances near the 737 nm wavelength range which will allow for efficient and resonantly enhanced excitation, control, and read out of SiV centers via a tapered optical fiber interface, to minimize optical probe heating effects. Furthermore, 2D structures realized in recently demonstrated diamond membranes [35] could be utilized to increase the thermal conductivity between an OMC device and its cold thermal bath, thus minimizing heating effects [34]. With the reduced heating expected from these new designs, we expect to detect signatures of the OMC's acoustic resonance in spin spectroscopy.

## 5. Conclusions

In summary, we have demonstrated high-Q diamond OMC devices with single SiV centers and measured a high-frequency (~6 GHz) mechanical mode with a linewidth of 13±1 kHz ($Q$~$4.4 \times 10^5$) at millikelvin temperatures, within an order of magnitude of state-of-the art linewidths in silicon OMCs [23]. The usage of phononic bandgap shields between the OMC and the bulk substrate could further improve the mechanical Q factors [23]. The pump power dependence of this linewidth allowed us to measure an optomechanical coupling rate of 43.5±0.2 kHz at the single photon and phonon level, a value comparable with previous diamond OMCs. We confirmed that single SiV centers can maintain good optical properties even in the nanocavity, and that their spins can be initialized and read out. From the pump power dependence of the OMC mechanical linewidth and spin lifetime measurements, we found that the coherence of the SiV spin and the OMC mechanical mode could be negatively impacted by laser-induced heating. The use of an efficient tapered diamond waveguide - tapered optical fiber interface [28], and of optical cavities resonant with the SiV ZPL [26] could greatly reduce optical probe powers and the associated heating. Based on the minimum measured mechanical linewidth ($\kappa = 13 \pm 1$ kHz), typical SiV spin decoherence rates ($\gamma$~100 Hz) [21] and the simulated spin-phonon coupling rate $g_{SP}/2\pi$ (1.65 MHz), our OMC system could reach the strong spin-phonon coupling regime where $g_{SP} > \kappa, \gamma$. With reduced optical heating effects, we envision that high Q-factor diamond OMCs with SiV centers will allow the investigation of spin-phonon interactions in highly engineered acoustic structures, and enable phonon-mediated interfaces between spins and other quantum platforms.

**Funding:**
NSF Engineering Research Center for Quantum Networks (EEC-1941583), NSF Science and Technology Center for Integrated Quantum Materials (NSF DMR-1231319), AFOSR (FA9550-20-1-01015 and FA9550-23-1-0333), ARO (W911NF1810432), ONR (N00014-20-1-2425), Harvard Quantum Initiative (HQI).

**Acknowledgment.**
The authors would like to thank M. Bhaskar, D. Assumpcao, and C. Knaut for the useful discussions. G. D. Joe was supported in part by the Natural Sciences and Research Council of Canada (NSERC). C. Chia was supported in part by Singapore's Agency for Science, Technology and Research (A*STAR). B. P. acknowledges financial support through a Horizon 2020 Marie Skłodowska-Curie Actions global fellowship (COHESiV, Project No. 840968) from the European Commission. K.K. acknowledges financial support from JSPS Overseas Research Fellowships (Project No. 202160592). S.M. acknowledges support from the IQIM Postdoctoral Fellowship. This work was performed in part at the Center for Nanoscale Systems (CNS), a member of the National Nanotechnology Infrastructure Network (NNIN), which is supported by the National Science Foundation award ECS-0335765. CNS is part of Harvard University.

**References**
[1] R. Manenti *et al.*, "Circuit quantum acoustodynamics with surface acoustic waves," *Nat. Commun.*, vol. 8, no. 1, Art. no. 1, Oct. 2017, doi: 10.1038/s41467-017-01063-9.
[2] D. A. Golter, T. Oo, M. Amezcua, I. Lekavicius, K. A. Stewart, and H. Wang, "Coupling a Surface Acoustic Wave to an Electron Spin in Diamond via a Dark State," *Phys. Rev. X*, vol. 6, no. 4, p. 041060, Dec. 2016, doi: 10.1103/PhysRevX.6.041060.
[3] S. Maity *et al.*, "Coherent acoustic control of a single silicon vacancy spin in diamond," *Nat. Commun.*, vol. 11, no. 1, Art. no. 1, Jan. 2020, doi: 10.1038/s41467-019-13822-x.
[4] M. Eichenfield, J. Chan, R. M. Camacho, K. J. Vahala, and O. Painter, "Optomechanical crystals," *Nature*, vol. 462, no. 7269, Art. no. 7269, Nov. 2009, doi: 10.1038/nature08524.
[5] J. Chan *et al.*, "Laser cooling of a nanomechanical oscillator into its quantum ground state," *Nature*, vol. 478, no. 7367, Art. no. 7367, Oct. 2011, doi: 10.1038/nature10461.
[6] A. H. Safavi-Naeini, S. Gröblacher, J. T. Hill, J. Chan, M. Aspelmeyer, and O. Painter, "Squeezed light from a silicon micromechanical resonator," *Nature*, vol. 500, no. 7461, Art. no. 7461, Aug. 2013, doi: 10.1038/nature12307.
[7] S. Meesala *et al.*, "Non-classical microwave-optical photon pair generation with a chip-scale transducer," arXiv.org. Accessed: Oct. 17, 2023. [Online]. Available: https://arxiv.org/abs/2303.17684v1
[8] P. K. Shandilya, D. P. Lake, M. J. Mitchell, D. D. Sukachev, and P. E. Barclay, "Optomechanical interface between telecom photons and spin quantum memory," *Nat. Phys.*, vol. 17, no. 12, Art. no. 12, Dec. 2021, doi: 10.1038/s41567-021-01364-3.
[9] S. Meesala *et al.*, "Strain engineering of the silicon-vacancy center in diamond," *Phys. Rev. B*, vol. 97, no. 20, p. 205444, May 2018, doi: 10.1103/PhysRevB.97.205444.
[10] M. Atatüre, D. Englund, N. Vamivakas, S.-Y. Lee, and J. Wrachtrup, "Material platforms for spin-based photonic quantum technologies," *Nat. Rev. Mater.*, vol. 3, no. 5, Art. no. 5, May 2018, doi: 10.1038/s41578-018-0008-9.
[11] A. M. Zaitsev, *Optical Properties of Diamond: A Data Handbook*. Springer Science & Business Media, 2013.
[12] Y. Tao, J. M. Boss, B. A. Moores, and C. L. Degen, "Single-crystal diamond nanomechanical resonators with quality factors exceeding one million," *Nat. Commun.*, vol. 5, no. 1, Art. no. 1, Apr. 2014, doi: 10.1038/ncomms4638.

[13] P. Ovartchaiyapong, K. W. Lee, B. A. Myers, and A. C. B. Jayich, "Dynamic strain-mediated coupling of a single diamond spin to a mechanical resonator," *Nat. Commun.*, vol. 5, no. 1, Art. no. 1, Jul. 2014, doi: 10.1038/ncomms5429.

[14] S. Meesala *et al.*, "Enhanced Strain Coupling of Nitrogen-Vacancy Spins to Nanoscale Diamond Cantilevers," *Phys. Rev. Appl.*, vol. 5, no. 3, p. 034010, Mar. 2016, doi: 10.1103/PhysRevApplied.5.034010.

[15] H. Chen *et al.*, "Engineering Electron–Phonon Coupling of Quantum Defects to a Semiconfocal Acoustic Resonator," *Nano Lett.*, vol. 19, no. 10, pp. 7021–7027, Oct. 2019, doi: 10.1021/acs.nanolett.9b02430.

[16] Y.-I. Sohn *et al.*, "Controlling the coherence of a diamond spin qubit through its strain environment," *Nat. Commun.*, vol. 9, no. 1, Art. no. 1, May 2018, doi: 10.1038/s41467-018-04340-3.

[17] S. Maity *et al.*, "Mechanical Control of a Single Nuclear Spin," *Phys. Rev. X*, vol. 12, no. 1, p. 011056, Mar. 2022, doi: 10.1103/PhysRevX.12.011056.

[18] A. Sipahigil *et al.*, "Indistinguishable Photons from Separated Silicon-Vacancy Centers in Diamond," *Phys. Rev. Lett.*, vol. 113, no. 11, p. 113602, Sep. 2014, doi: 10.1103/PhysRevLett.113.113602.

[19] M. J. Burek *et al.*, "Diamond optomechanical crystals," *Optica*, vol. 3, no. 12, pp. 1404–1411, Dec. 2016, doi: 10.1364/OPTICA.3.001404.

[20] J. V. Cady *et al.*, "Diamond optomechanical crystals with embedded nitrogen-vacancy centers," *Quantum Sci. Technol.*, vol. 4, no. 2, p. 024009, Mar. 2019, doi: 10.1088/2058-9565/ab043e.

[21] D. D. Sukachev *et al.*, "Silicon-Vacancy Spin Qubit in Diamond: A Quantum Memory Exceeding 10 ms with Single-Shot State Readout," *Phys. Rev. Lett.*, vol. 119, no. 22, p. 223602, Nov. 2017, doi: 10.1103/PhysRevLett.119.223602.

[22] H. A. Atikian *et al.*, "Freestanding nanostructures via reactive ion beam angled etching," *APL Photonics*, vol. 2, no. 5, p. 051301, May 2017, doi: 10.1063/1.4982603.

[23] G. S. MacCabe *et al.*, "Nano-acoustic resonator with ultralong phonon lifetime," *Science*, vol. 370, no. 6518, pp. 840–843, Nov. 2020, doi: 10.1126/science.abc7312.

[24] R. Stockill *et al.*, "Gallium Phosphide as a Piezoelectric Platform for Quantum Optomechanics," *Phys. Rev. Lett.*, vol. 123, no. 16, p. 163602, Oct. 2019, doi: 10.1103/PhysRevLett.123.163602.

[25] C. Chia, B. Machielse, A. Shams-Ansari, and M. Lončar, "Development of hard masks for reactive ion beam angled etching of diamond," *Opt. Express*, vol. 30, no. 9, pp. 14189–14201, Apr. 2022, doi: 10.1364/OE.452826.

[26] C. T. Nguyen *et al.*, "An integrated nanophotonic quantum register based on silicon-vacancy spins in diamond," *Phys. Rev. B*, vol. 100, no. 16, p. 165428, Oct. 2019, doi: 10.1103/PhysRevB.100.165428.

[27] R. E. Evans *et al.*, "Photon-mediated interactions between quantum emitters in a diamond nanocavity," *Science*, vol. 362, no. 6415, pp. 662–665, Nov. 2018, doi: 10.1126/science.aau4691.

[28] M. J. Burek *et al.*, "Fiber-Coupled Diamond Quantum Nanophotonic Interface," *Phys. Rev. Appl.*, vol. 8, no. 2, p. 024026, Aug. 2017, doi: 10.1103/PhysRevApplied.8.024026.

[29] J. D. Cohen *et al.*, "Phonon counting and intensity interferometry of a nanomechanical resonator," *Nature*, vol. 520, no. 7548, Art. no. 7548, Apr. 2015, doi: 10.1038/nature14349.

[30] M. Aspelmeyer, T. J. Kippenberg, and F. Marquardt, "Cavity optomechanics," *Rev. Mod. Phys.*, vol. 86, no. 4, pp. 1391–1452, Dec. 2014, doi: 10.1103/RevModPhys.86.1391.

[31] T. P. M. Alegre, A. H. Safavi-Naeini, M. Winger, and O. Painter, "Quasi-two-dimensional optomechanical crystals with a complete phononic bandgap," *Opt. Express*, vol. 19, no. 6, p. 5658, 2011.

[32] T. J. Kippenberg and K. J. Vahala, "Cavity Optomechanics: Back-Action at the Mesoscale," *Science*, vol. 321, no. 5893, pp. 1172–1176, Aug. 2008, doi: 10.1126/science.1156032.

[33] W. Jiang *et al.*, "Lithium niobate piezo-optomechanical crystals," *Optica*, vol. 6, no. 7, pp. 845–853, Jul. 2019, doi: 10.1364/OPTICA.6.000845.

[34] H. Ren *et al.*, "Two-dimensional optomechanical crystal cavity with high quantum cooperativity," *Nat. Commun.*, vol. 11, no. 1, Art. no. 1, Jul. 2020, doi: 10.1038/s41467-020-17182-9.

[35] X. Guo *et al.*, "Tunable and Transferable Diamond Membranes for Integrated Quantum Technologies," *Nano Lett.*, vol. 21, no. 24, pp. 10392–10399, Dec. 2021, doi: 10.1021/acs.nanolett.1c03703.

# High Q-factor diamond optomechanical resonators with silicon vacancy centers at millikelvin temperatures: supplementary material

**Design details for OMCs**

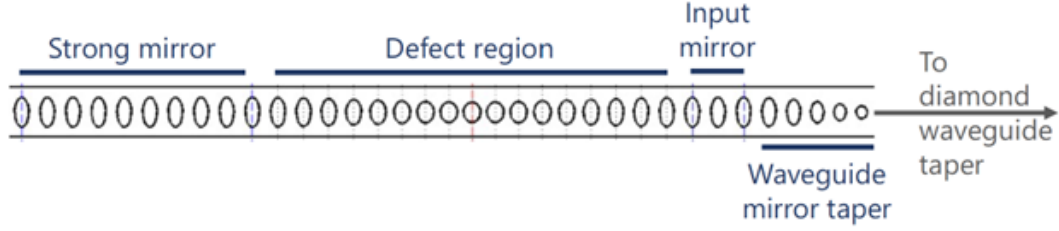

Fig. S1. In-plane geometry of our OMC designed for single-ended coupling, showing an engineered input mirror consisting of a waveguide mirror taper and reduced number of mirror holes.

To form simultaneous photonic and phononic "defect" cavity modes, we locally vary the lattice constant and eccentricity of the air-holes with two additional parameters d and b defining the defect unit cell parameters as:

$$a_{def}(n) = a\left(1 - d(f(n))\right) \tag{S1}$$

$$h_{x,def}(n) = h_x(1 - d(f(n)))^{1-b} \tag{S2}$$

$$h_{y,def}(n) = h_y(1 - d(f(n)))^{1+b} \tag{S3}$$

Where $f(n) = 1 - 3n^2 + 2n^3$, $n = (k-1)/n_{def}$ is an index that runs from 0 to 1, $n_{def}$ is the number of defect holes from the central defect hole to the mirror segment, and k is 1 for the central defect hole and $n_{def}$ for the defect hole adjacent to the mirror segment. For our optimized design, we used the nominal mirror unit cell parameters $(a, w, h_x, h_y) = (528, 936, 197, 578)$ nm and cavity defect tapering parameters $(d, b) = (0.19258, 2.2761)$.

We also tapered the waveguide OMC-mirror interface to minimize scattering losses. This waveguide mirror taper (WMT) consists of a series of air holes that gradually taper in size starting from small holes at the input end to larger holes resembling the OMC mirror. The WMT unit cells are defined by tapering the difference between the end WMT unit cell and the OMC mirror cell by

$$a_{WMT}(m) = a_{end} + (a - a_{end})f(m) \tag{S4}$$

$$h_{x,WMT}(m) = h_{x,end} + (h_x - h_{x,end})f(m) \tag{S5}$$

$$h_{y,WMT}(m) = h_{y,end} + (h_y - h_{y,end})f(m) \tag{S6}$$

Where $m = (k-1)/n_{WMT}$ runs from 0 to 1, $n_{WMT}$ is the number of holes in the WMT, and k indexes the holes in the WMT starting from the OMC mirror cell. The taper function is $f(m) = 1 - 3m^2 + 2m^3$. We set $n_{WMT} = 5$ as a compromise between efficiency and waveguide coupling rate.

Finally, we reduced the number of input mirror cells relative to the strong mirror to increase the coupling rate between the defect region of the OMC and the WMT. We aim for the critical coupling condition where the waveguide cavity coupling rate $\kappa_e$ is equal to the intrinsic cavity loss rate $\kappa_i$, where the on-resonance reflection contrast is maximum. FDTD modeling which attempts to account for fabrication induced losses estimates that three input mirror cells are necessary to reach this condition [1]. To take the uncertainty in fabrication induced losses into account, OMC devices with different numbers of input mirror cells are fabricated.

**Simulation of Diamond OMCs**

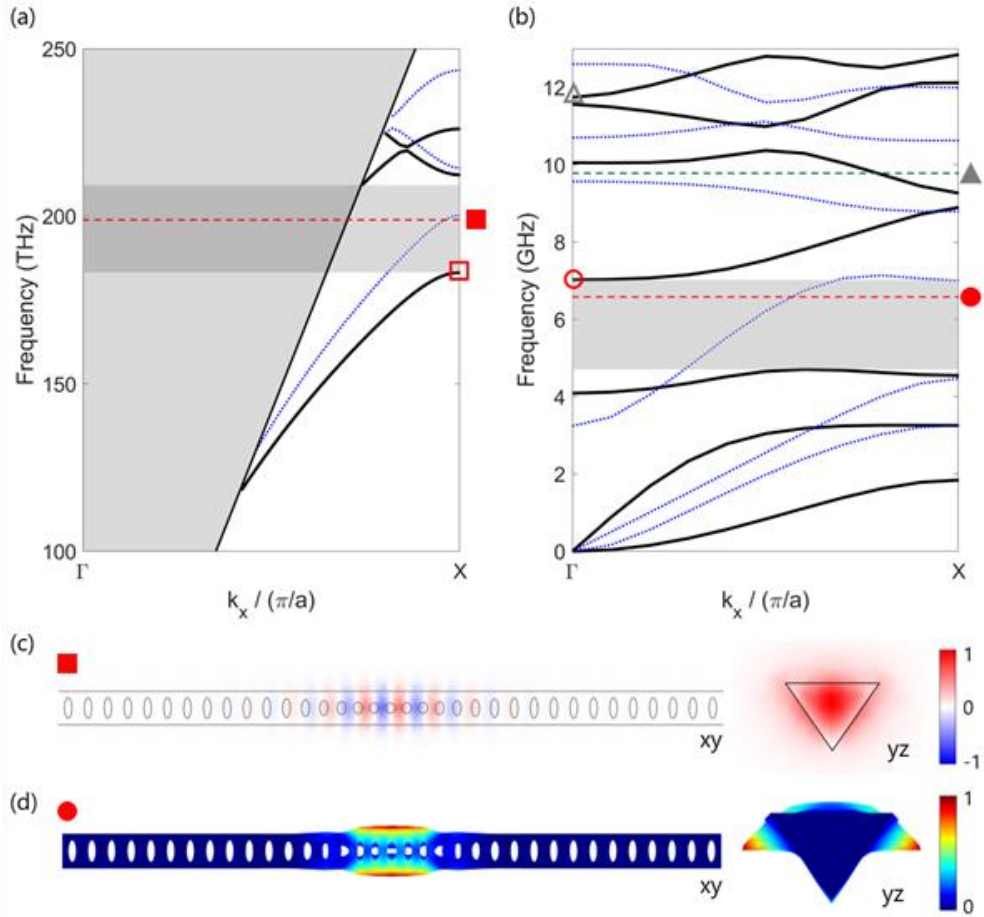

Fig. S2. Optimized OMC unit cell and localized modes. (a) Optical bandstructure of the unit cell. Solid black and dotted blue curves beneath the light cone (gray shaded area) represent TE-like and TM-like modes, with the TE

quasi-bandgap represented by the light gray rectangle. The red hollow square denotes the TE-like guided mode at the X-point, while the red solid square denotes the localized TE-like mode in the OMC. (b) Acoustic bandstructure of the unit cell. Solid black and dotted blue lines represent modes of even and odd symmetries about the yz-plane. The red hollow circle and green hollow triangle denote guided acoustic flexural and compressive modes, while the red solid circle and green solid triangle denote localized OMC acoustic flexural and compressive modes respectively. (c) Simulated electric field profile (y-component) of the localized TE-like mode in the OMC. (d) Simulated displacement field profile of the localized acoustic flexural mode in the OMC.

Optical and acoustic bandstructure simulations for single unit cells were performed as sweeps of eigenmode simulations. The yz-planes at both ends of the unit cell were specified to have Bloch symmetry with wavevector k to account for periodicity across the lattice constant a. The wavevector k was swept across multiple values between the Γ- and X-points inclusive, and an eigenmode simulation was performed at each value of k. To account for modes of all symmetries, the sweep was repeated for all possible symmetries. The optical and mechanical modes of diamond OMCs are simulated using the finite-difference time-domain (FDTD) method (Lumerical) and finite element method (FEM) (COMSOL). We focused on quasi-TE optical and symmetric mechanical modes since they maximize the overlap between optical and mechanical displacement/stress fields [2]. In our convention, TE-like modes have odd vector symmetry with respect to reflection across the $y = 0$ longitudinal symmetry plane of the nanobeam [see inset of Fig. 1(b) for axis definitions].

From its optical bandstructure, the mirror unit cell supports a quasi-bandgap from 183.3 to 209.3 THz, with the TE-like guided mode at the X-point at 183.3 THz situated beneath the quasi-bandgap [Fig. S2(a)]. The defect region, with hole dimensions decreasing in eccentricity from the mirror to the central defect cell, raises the mode frequency into the quasi-bandgap above it. The localized TE-like mode supported by the triangular OMC has a resonance wavelength of 1507 nm (198.9 THz) [Fig. S2(c)]. Additionally, the acoustic bandstructure of the mirror unit cell shows a symmetry bandgap from 4.71 to 7.02 GHz, with the flexural and compressive Γ-point guided modes at 7.02 and 11.74 GHz respectively [Fig. S2(b)]. The defect region leads to lowering of acoustic mode frequencies, and the localized acoustic flexural mode of the OMC has a mode frequency of 6.58 GHz, within the symmetry bandgap [Fig. S2(d)]. While the compressive mode is also well localized within the defect region, we do not study it in this work.

The total optomechanical coupling rate $g_{OM}$ is calculated as the sum of cavity frequency ($\omega_{opt}/2\pi = f_{opt}$) shifts from moving boundary (MB) and photoelastic effects (PE), multiplied by the zero-point fluctuation amplitude:

$$g_{OM} = \left(\frac{d\omega_{opt}}{dx}\bigg|_{MB} + \frac{d\omega_{opt}}{dx}\bigg|_{PE}\right) x_{zpf} \tag{S7}$$

$g_{OM}$ is maximized when the overlap integrals between the electric field and displacement field (MB) or strain field (PE) are maximized. The spin-phonon coupling rate $g_{SP}$ at any point in the geometry was calculated using the strain tensor at that point arising from the mechanical flapping mode. The optimized design was obtained using a simplex search algorithm that maximizes a product of $g_{SP}$ for an SiV center spin implanted at the optimal position of the OMC mechanical flapping mode, $g_{OM}$ between the mechanical flapping mode and optical quasi-TE mode, total optical quality factor $Q_{opt}$, and optical transmission T between the OMC and the diamond waveguide [1]. We note that while $g_{SP}$ is maximized at the surface, we optimized for the value at a target SiV depth of 20 nm below the diamond surface - a tradeoff between SiV stability and the large $g_{SP}$ needed to observe coherent spin-phonon interactions. Taking this into account, our optimized geometry (Fig. 1) with an optimally placed SiV center can achieve $g_{SP}$ approaching 1.65 MHz (assuming zero static crystal strain at the location of SiV). $Q_{opt}$ was capped with a cutoff at $5\times10^6$ to prevent artificially high but experimentally unrealizable Qs from excessively weighing the fitness function and skewing the search. Finally, a large T maximizes coupling between the OMC and diamond waveguide thus enabling efficient excitation with a tapered optical fiber. [3] For each iteration of the algorithm, the in-plane geometry was parameterized by the bandgap unit cell geometry *(a, w, h_x, h_y)*, as well as two defect parameters *(d, b)* which control the tapering of air hole perforations in the cavity region. Such a taper aims to localize the optical and mechanical modes in the cavity region by raising and lowering the optical and mechanical mode frequencies into their respective bandgaps.

**Fabrication of Optomechanical Crystals**

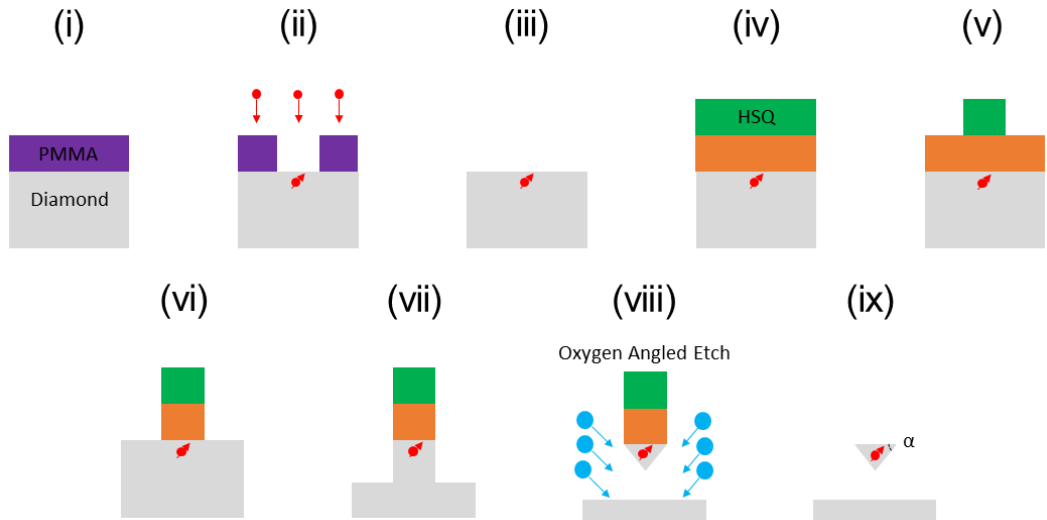

Fig.S3. Fabrication of OMC-SiV system. (i) Spin coating of PMMA, (ii) electron beam lithography, development of PMMA template, and silicon ion implantation, (iii) removal of PMMA, (iv) sputtering of Nb followed by spin coating of HSQ, (v) electron beam lithography and development of HSQ, (vi) anisotropic etching of Nb, (vii) anisotropic oxygen etching of diamond, (viii) angled oxygen etching of diamond, (ix) mask removal. α defines the etch angle.

The fabrication process for OMCs with SiV centers started with etching of alignment markers into electronic-grade single-crystal diamond (EL SC, [N] < 5 ppb, Element Six), followed by implantation of silicon ions through an electron beam resist mask, and then subsequent annealing to generate SiV centers [4]. A depth of 20 nm was chosen for SiV, this depth corresponds to a beam energy of 27.5 keV as calculated using SRIM [5]. The implantation dose was $1.25 \times 10^{12}$/cm$^2$ aiming for approximately 3 to 4 SiV centers per implantation aperture, which were positioned between the central defect hole in the OMC and the holes adjacent to it on both sides. The fabrication process continued with mask definition, top-down etching, angled-etching, and lastly mask removal and critical point drying (Fig. S3) [6]. The mask used was a 1 µm-thick layer of HSQ, on top of a 200 nm-thick layer of niobium (Nb) deposited via DC magnetron sputtering (AJA International). Nb was chosen due to its low erosion rate in oxygen RIBAE [7], and its low extent of mask redeposition which enabled smooth triangular cross-sections to be realized. The Nb thickness was chosen to complement the HSQ mask thickness in terms of withstanding mask erosion and providing shadowing of holes in the OMC during the angled etching process. Finally, it provides good contrast with HSQ and diamond under electron beam exposure. As the final OMCs need to be aligned to the previously implanted apertures, clear inspection and precise location of the alignment markers in the SEM is needed prior to electron beam lithography. Nb, with its higher atomic number, provides sufficient secondary electron scattering to the SEM detector and hence also provides clear imaging of the marker for aligning the mask to. The OMC mask geometry was aligned to the alignment markers during electron beam lithography. As the implantation apertures were also aligned to the same markers, this ensures that the SiV spots would be aligned to the central defect of the OMCs. After development, but prior to the top-down etch of diamond, the mask pattern was transferred to Nb using argon/chlorine mixture (flow rates 25/40 sccm, pressure 8 milliTorr, RF power 250 W, ICP power 400 W). After angled etching, the Nb and HSQ were removed in a 1:1 mixture of 49% hydrofluoric acid and 60% nitric acid, the devices were then cleaned with piranha mixture and finally critical point dried with carbon dioxide.

## Telecom quenching of SiV fluorescence

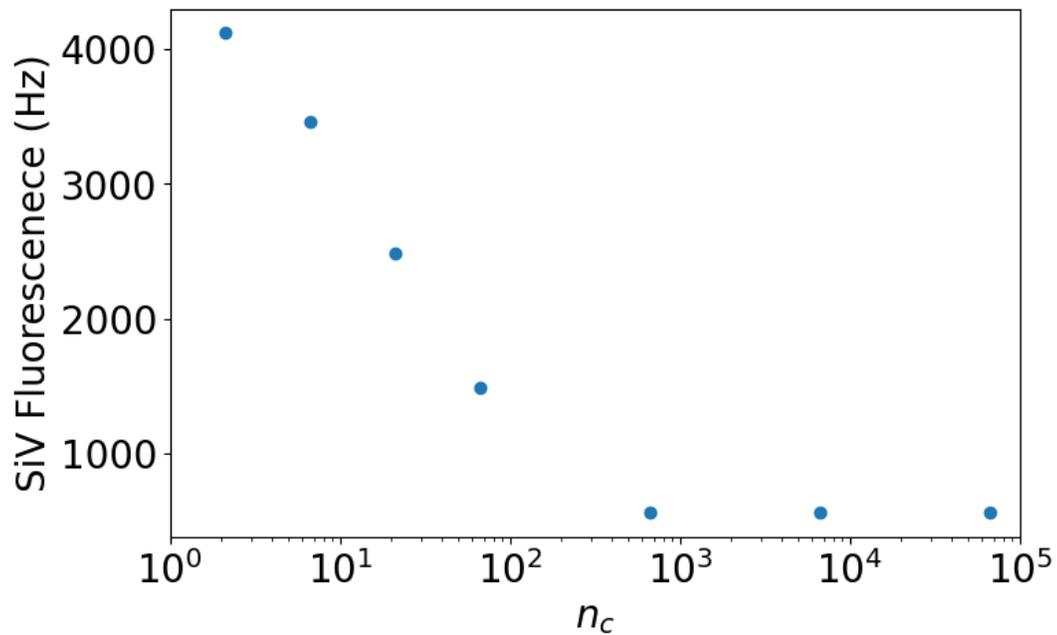

Fig. S4. Quenching of SiV fluorescence when telecom light at detuning of $\Delta = \kappa/2$ from the optical resonance is incident on the OMC.

An unexpected observation during this work was that SiV fluorescence measured through the confocal microscope can be strongly quenched by telecom wavelength light sent through the tapered waveguide - tapered optical fiber interface. A telecom laser power dependence curve of this quenching behavior is shown in Fig. S4.

**Pump-Probe Measurements of SiV Spin Lifetime**

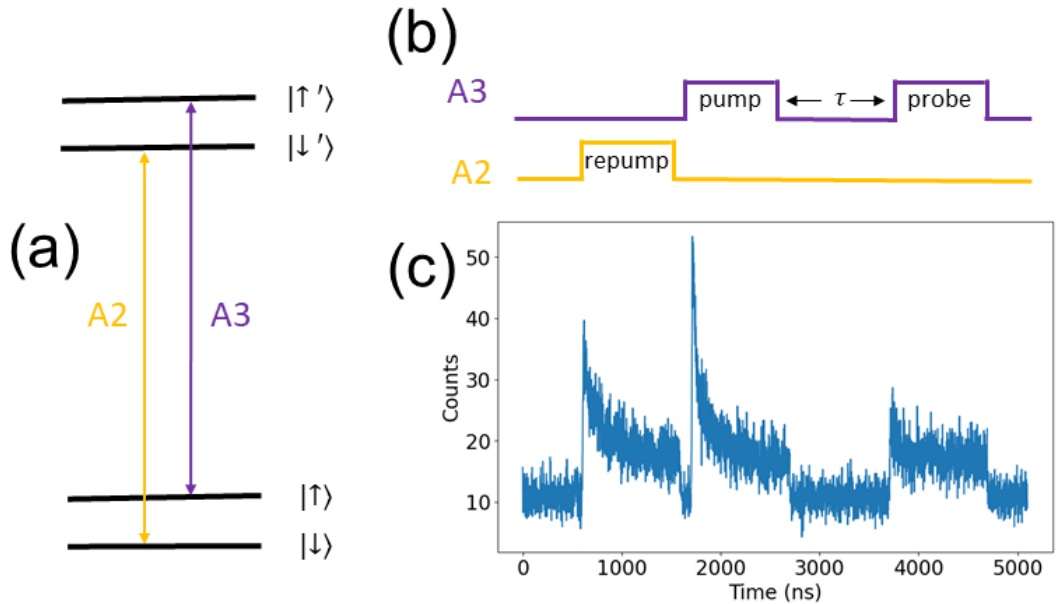

Fig. S5. Pump-probe measurement of SiV spin lifetime. (a) Simplified level structure of the SiV. A2 and A3 are the spin-conserving transitions. (b) Pulse sequence for spin T1 measurement (c) Time-resolved PLE signal for this sequence.

A two-laser pulse sequence is used to measure the spin lifetime of the SiV. A repump A2 pulse is used to pump any residual population into the $|\uparrow\rangle$ state, then a pump A3 pulse initializes all population into the $|\downarrow\rangle$ state, followed by an inter-pulse time delay $\tau$ and a probe A3 pulse which reads out the population in $|\uparrow\rangle$. Sweeping $\tau$, we can infer the SiV spin T1 (Fig. 4(c) inset).

Note that at the millikelvin temperatures used in this work, cw-PLE is not possible with a single laser due to the long spin T1 lifetimes relative to 4 K temperatures. Despite the short T1 times observed in this work relative to other reports at millikelvin temperatures [8], a two-laser approach where one laser is stationary and acting as a repump laser while the other is scanned was still necessary to observe the PLE spectra in Fig. 4(c).

## Integrated mechanical spectra

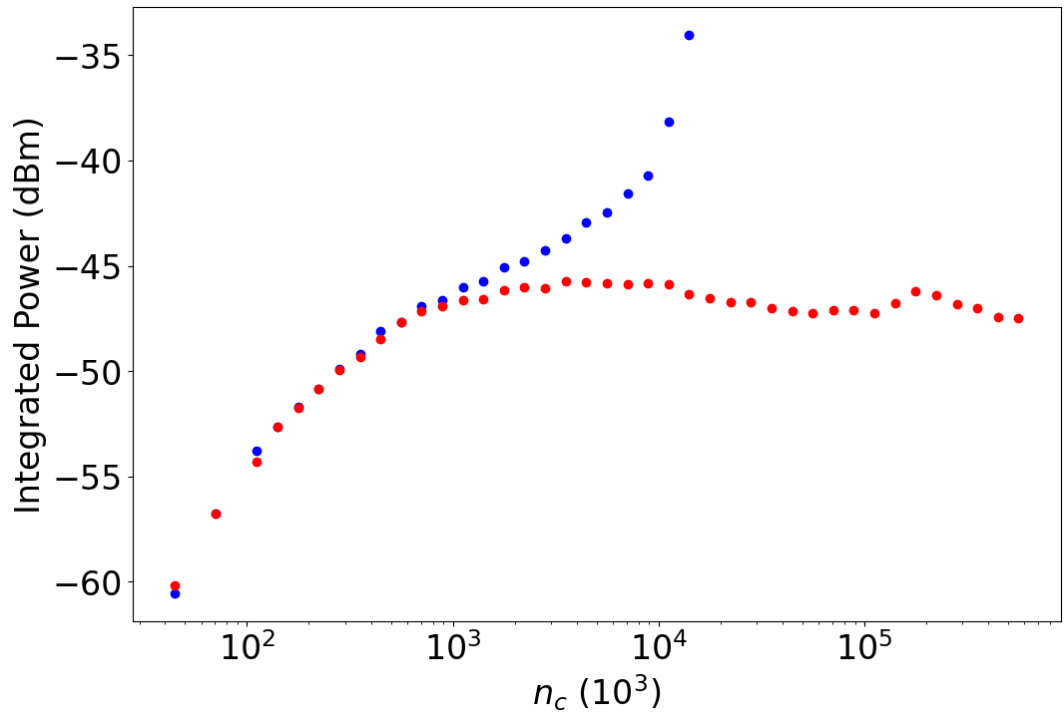

Fig. S6. Integrated mechanical PSD spectra vs. intracavity photon number $n_c$. with the laser detuning set to $\kappa/2$ (blue dots) and $-\kappa/2$ (red dots). The onset of mechanical lasing is evident on the blue sideband and cooling on the red sideband at high $n_c$ is visible. If carefully calibrated to a modulator generated sideband of known power, this measurement can be used to infer the phonon occupation of the cavity mode, this will be a topic of future study.

## Optical setup for millikelvin optomechanical spectroscopy

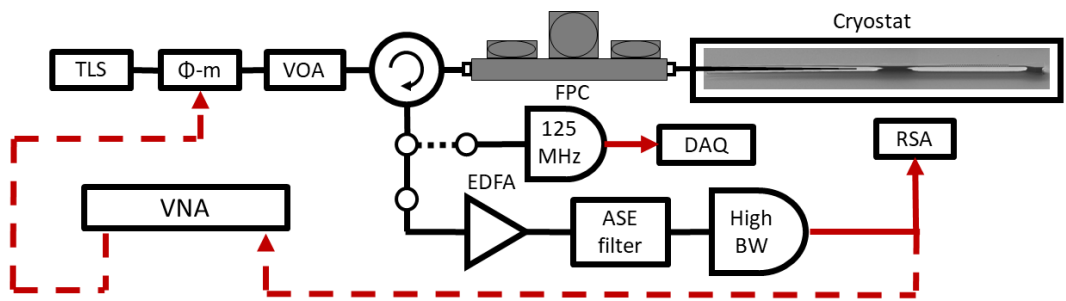

Fig. S7. Detailed optical setup used for optical and mechanical spectroscopy of OMCs at millikelvin temperatures. TLS: tunable laser source, Φ-m: electro-optic phase modulator, VOA: variable optical attenuator, FPC: fiber polarization controller, DAQ: data acquisition device, EDFA: Erbium doped fiber amplifier, ASE filter: amplified spontaneous emission noise filter, RSA: real-time spectrum analyzer, VNA: vector network analyzer. Measured device coupling efficiencies on this setup are ~44%.

For optomechanical spectroscopy at millikelvin temperatures the setup pictured in Fig. S6 was used. A tunable laser source (TLS, Santec TSL 710), fiber circulator (Thorlabs 6015-3), and a high gain low bandwidth photodetector (New Focus 1811, 125 MHz) was used to measure OMC reflection spectra. Laser input power was controlled with a variable optical attenuator (VOA, EXFO FA-3150), and polarization was set with a fiber polarization controller (FPC). Mechanical spectroscopy was performed by first amplifying the reflected optical signal with an erbium doped fiber amplifier (EDFA, PriTel LNTFA-20-L-10), followed by a filter for amplified spontaneous emission noise (ASE filter, JDS Uniphase TB9) and then measured with a high bandwidth photodetector (New Focus 1551-B, 12 GHz bandwidth) connected to a real-time spectrum analyzer (RSA, Tektronix RSA 5126A). The phase response of the optical cavity was measured by driving a phase modulator (Φ-m, EOSpace Inc.) with a vector network analyzer (VNA, Agilent E8364B) and detecting the RF beat note on a higher bandwidth photodetector (New Focus 1014, 45 GHz bandwidth). System losses were calibrated during setup, allowing us to know the input power $P_{in}$ at the device. The intracavity photon number is related to this $P_{in}$ by

$$n_c = \frac{P_{in}}{\hbar \omega_l} \frac{\kappa_e}{\Delta^2 + (\kappa/2)^2} \qquad (S8)$$

Where $\omega_l$ is the laser frequency.

**References**

[1] "Quantum Optomechanics with Color Centers in Diamond." Accessed: Oct. 01, 2023. [Online]. Available: https://dash.harvard.edu/handle/1/37368314

[2] M. Eichenfield, J. Chan, A. H. Safavi-Naeini, K. J. Vahala, and O. Painter, "Modeling dispersive coupling and losses of localized optical and mechanical modes in optomechanical crystals," *Opt. Express*, vol. 17, no. 22, pp. 20078–20098, Oct. 2009, doi: 10.1364/OE.17.020078.

[3] M. J. Burek *et al.*, "Fiber-Coupled Diamond Quantum Nanophotonic Interface," *Phys. Rev. Appl.*, vol. 8, no. 2, p. 024026, Aug. 2017, doi: 10.1103/PhysRevApplied.8.024026.

[4] C. T. Nguyen *et al.*, "An integrated nanophotonic quantum register based on silicon-vacancy spins in diamond," *Phys. Rev. B*, vol. 100, no. 16, p. 165428, Oct. 2019, doi: 10.1103/PhysRevB.100.165428.

[5] "SRIM & TRIM. v. 2018b." [Online]. Available: https://www.mathworks.com/products/matlab.html

[6] C. Chia, B. Machielse, A. Shams-Ansari, and M. Lončar, "Development of hard masks for reactive ion beam angled etching of diamond," *Opt. Express*, vol. 30, no. 9, pp. 14189–14201, Apr. 2022, doi: 10.1364/OE.452826.

[7] H. A. Atikian *et al.*, "Diamond mirrors for high-power continuous-wave lasers," *Nat. Commun.*, vol. 13, no. 1, Art. no. 1, May 2022, doi: 10.1038/s41467-022-30335-2.

[8] D. D. Sukachev *et al.*, "Silicon-Vacancy Spin Qubit in Diamond: A Quantum Memory Exceeding 10 ms with Single-Shot State Readout," *Phys. Rev. Lett.*, vol. 119, no. 22, p. 223602, Nov. 2017, doi: 10.1103/PhysRevLett.119.223602.